\documentclass[fleqn,10pt]{wlscirep}
\usepackage[utf8]{inputenc}
\usepackage[T1]{fontenc}

\usepackage[autostyle=true]{csquotes} 

\usepackage{subfig}
\usepackage{physics}
\usepackage[acronym]{glossaries}
\newacronym{ssfm}{SSFM}{Split Step Fourier Method}
\newacronym{bec}{BEC}{Bose-Einstein condensates}
\newacronym{gpe}{GPE}{Gross-Pitaevskii equation}
\newacronym{nlse}{NLSE}{Nonlinear Schrödinger equation}
\newacronym{npse}{NPSE}{Non-Polynomial Schrödinger equation}
\newacronym{pbc}{PBC}{Periodic Boundary Conditions}

\newcommand{\cblue}{\color{black}}
\newcommand{\cblueA}{\color{black}}
\newcommand{\cplus}{\color{black}}

\title{Atomic soliton transmission and induced collapse in scattering from a narrow barrier}

\author[1, 2, *]{Francesco Lorenzi}
\author[1, 2, 3, 4]{Luca Salasnich}
\affil[1]{Dipartimento di Fisica e Astronomia "Galileo Galilei", 
Universit\`a di Padova, Via Marzolo 8, 35131 Padova, Italy}
\affil[2]{Istituto Nazionale di Fisica Nucleare (INFN), Sezione di Padova, via Marzolo 8, 35131 Padova, Italy}
\affil[3]{Padua Quantum Technology Research Center, Universit\`a di Padova, Via Gradenigo 6/A, 35131 Padova, Italy}
\affil[4]{Istituto Nazionale di Ottica (INO) del Consiglio Nazionale delle Ricerche (CNR), via Nello Carrara 1, 50019 Sesto Fiorentino, Italy}

\affil[*]{francesco.lorenzi.2@phd.unipd.it}

\begin{abstract}
We report systematic numerical simulations of the collision of a bright matter-wave soliton made of Bose-condensed alkali-metal atoms through a narrow potential barrier by using the three-dimensional Gross-Pitaevskii equation.
In this way, we determine how the transmission coefficient depends on the soliton impact velocity and the barrier height. Quite remarkably, we also obtain the regions of parameters where there is the collapse of the bright soliton induced by the collision.
 We compare these three-dimensional results with the ones obtained by three different one-dimensional nonlinear Schr\"odinger equations. We find that a specifically modified nonpolynomial Schr\"odinger equation {\cblueA is able to accurately assess the transmission coefficient even in a region in which the usual nonpolynomial Schr\"odinger equation collapses}. In particular, this simplified but very effective one-dimensional model takes into account the transverse width dynamics of the soliton with an ordinary differential equation coupled to the partial differential equation of the axial wave function of the Bose-Einstein condensate.
\end{abstract}
\begin{document}

\flushbottom
\maketitle

\thispagestyle{empty}

Localized soliton-like structures, known as bright matter-wave solitons, can be generated in Bose-Einstein Condensates (BEC) with attractive interatomic interactions. Since the first experimental realization of such structure about two decades ago \cite{khaykovich-formation-2002}, dynamics of matter-wave bright solitons in an attractive BEC have been intensely studied, both at the quantum level and using the Gross-Pitaevskii equation (GPE). Current experimental capabilities offer an unprecedented opportunity to test many-body theories in an ultracold Bose gas, and matter-wave solitons are an excellent target for the predictions \cite{weiss-creation-2009,streltsov-scattering-2009}. Moreover, several technological applications are based on the possibility of generating and manipulating such kind of coherent structures: some of them are interferometry \cite{helm-sagnac-2015} even beyond the quantum limit, and quantum-enhanced metrology \cite{dunningham-sub-shot-noise-limited-2004,dunningham-using-2006}.
The remarkable analogy of models based on the 3D-GPE with the equations of motion occurring in optics with Kerr media allowed to study common aspects on the same ground, such as the Hong-Ou-Mandel experiment \cite{sun-mean-field-2014}. 

In a typical setup of an atomic interferometer using solitons, a matter-wave soliton is prepared in a quasi-1D trap, i.e. a confining potential made of a strong radial component and a weak or absent axial component. This setup experimentally allows the creation of cigar-shaped condensates in the case of repulsive interparticle interaction and in the noninteracting case. By using attractive interparticle interactions that are obtainable for example by using Feshbach resonances, one can generate a matter-wave soliton. In the latter case, solitons loaded into quasi-1D traps have typical axial widths that are comparable to the radial potential characteristic length \cite{}. To achieve interference, the soliton is set into motion by phase imprinting, and it collides with a narrow potential barrier set by a narrow laser beam, acting analogously to an optical beam splitter, designed to be able to split the number of atoms into two even solitonic packets. The resulting two solitons are then recombined in a later stage, through the same barrier. After the first splitting, split solitons may achieve a differential phase shift, thus allowing the observation of interference in the recombined packet.  

From the theoretical point of view, the study of quantum matter-wave solitons was carried out mostly in 1D, where the many-body wave function is well known to have an exact solution by Bethe ansatz \cite{lieb-exact-1963}. This procedure relies on having very strong radial confinement, and it is not sensible to the 3D dynamics, thus losing details about the transverse degrees of freedom that are especially important near the point of GPE collapse in which the reduction of the size of the condensate brings it to a regime in which other interaction effects start to be non-negligible, like three-body interactions causing depletion from the trap \cite{cornish-formation-2006}.

The stability of solitons in quasi-1D harmonic traps is highly nontrivial, as the cubic GPE has a critical dimension {\cplus equal to } $2$. Instability is in the form of a collapse, also known in the mathematical literature as nonlinear blow-up \cite{sulem-nonlinear-2007}. In one dimension, the collapse is prevented by the Vlasov-Petrishev-Talanov theorem \cite{sulem-nonlinear-2007}. GPE collapse due to an arbitrary attractive interaction potential can be triggered in this context by loading into the trap a suitably high number of particles \cite{berge-wave-1998}. Moreover, the radial anisotropy of the trap can play a role in the critical number of particles for collapse \cite{mazzarella-collapse-2009,gammal-critical-2002}.
It is fundamental to remark that, even with a purely 1D model, by adding a barrier-like external potential to the 1D-GPE the problem becomes non-integrable, and requires approximate methods to be tackled.

The 3D-GPE dynamics represents a useful tool not only as an approximation of the full quantum dynamics but also as providing signatures of soliton entanglement across the barrier \cite{gertjerenken-generating-2013}. In fact, the discontinuity in the reflection coefficient indicates the possibility of creating ``Schrödinger cat" states, exploring quantum entanglement phenomena \cite{gertjerenken-generating-2013}.
The 3D-GPE model predicts a peculiar behavior of the transmission coefficient with the barrier: at low values of the velocity and the barrier height it is a discontinuous function of the parameters.
This aspect has been investigated \cite{gertjerenken-generating-2013} and is frequently referred to as the particle behavior of the impinging soliton.

Various dimensional reduction schemes for the 3D-GPE have been proposed, 
\cite{weiss-creation-2009,khaykovich-deviation-2006, salasnich-condensate-2002,salasnich-effective-2002}.
In this work, we compare three schemes of dimensional reduction with full 3D simulations. 1D effective equations have a great computational advantage in the description of the dynamics and are routinely used in studies of atomic interferometers.
The simplest one is the 1D-GPE, obtained by imposing a fixed transverse wave function as the lowest energy eigenstate of the transverse harmonic potential.
An improved model is called nonpolynomial Schr\"odinger equation (NPSE) \cite{salasnich-effective-2002}, which is based on assuming the transverse width of the trial as a variational parameter and obtaining the equation of motion as Euler-Lagrange (EL) equations. Furthermore, in the original NPSE formulation, derivatives of the transverse width parameter present in the Lagrangian are neglected, and the corresponding equation is algebraic. So we also consider the non-approximated version of the NPSE, which we call NPSE+ for brevity.
Previous work \cite{cuevas-interactions-2013} highlighted the behavior of the transmission coefficient and the barrier-induced collapse with the 3D-GPE and the NPSE, studying regions in the barrier height vs. number of particles plane. %
We investigate the differences in the collision process varying the velocity, focusing on the transmission coefficient and the onset of barrier-induced collapse.

\subsection*{Gross-Pitaevskii equations}
  The model is based on the Hartree approximation for bosons \cite{lieb-exact-1963}, using which it is possible to derive the following Lagrangian, called Gross-Pitaevskii Lagrangian, for the field $\psi({\bf r},t)$, representing the wave function of the Hartree product state, with all the particles in the same single-particle quantum state,
  \begin{equation}\label{eq:3dlagrangian}
      \mathcal{L}= \int \dd[3]{\mathbf{r}}  \ \psi^* \left[i \hbar \frac{\partial}{\partial t}+\frac{\hbar^2}{2 m} \nabla^2-U-\frac{g}{2}(N-1)|\psi|^2 \right] \psi,
  \end{equation}
  where $U$ is the external potential, $N$ is the number of particles, and $g$ is the contact potential, which can be linked to the s-wave scattering length $a_s$ with the expression 
  \begin{equation}
    g = \frac{4\pi \hbar^2 a_s}{m}.
  \end{equation}
  The associated {\cplus EL} equation is the 3D-GPE:
  \begin{equation}\label{eq:3dgpe}
      i\hbar \dfrac{\partial}{\partial t}\psi = \left[-\dfrac{\hbar^2}{2m} \nabla^2 + U  + g(N-1)|\psi|^2\right]\psi.
  \end{equation}

  Standard dimensional reduction of the 3D-GPE in a tight transverse harmonic potential relies on the assumption that the transverse degree of freedom of the wave function is frozen to the ground state of the harmonic potential, as we will briefly review now.
  Let the external potential be written as
  \begin{equation}
      U(x, y, z) = \frac{1}{2}m\omega_\perp^2 (y^2+z^2) + V(x),
  \end{equation}
  where $\omega_\perp$ is the (isotropic) strength of the potential, and $V$ is the axial part of the potential. The role of anisotropy on the transverse potential was studied in \cite{gammal-critical-2002, mazzarella-collapse-2009}. The strength $\omega_\perp$ naturally sets a characteristic length scale $l_\perp = \sqrt{\hbar/(m\omega_\perp)}$.
  Let us assume that the wave function is composed of a constant Gaussian transverse part $\phi$, which is the ground state of the transverse harmonic potential, and a time-varying axial component $f$ as 
  \begin{equation}\label{eq:1dansats}
      \psi(\mathbf{r}, t) = f(x, t) \phi(y, z),
  \end{equation}
  where
  \begin{equation}
      \phi(y, z) = \dfrac{1}{\sqrt{\pi} l_\perp} \exp\left[-\frac{y^2+z^2}{2l_\perp^2}\right].
  \end{equation}
  This is physically justified when the interaction energy is much smaller than the energy difference between the first excited state and the ground state of the transverse potential.
  Inserting this ansatz into Eq.~(\ref{eq:3dgpe}), and integrating along the transverse coordinates, one obtains the corresponding wave equation, called the 1D-GPE:
  \begin{equation}\label{eq:gpe}
    i\hbar \dfrac{\partial}{\partial t}f = \left[-\dfrac{\hbar^2}{2m}\dfrac{\partial^2}{\partial x^2} + V(x) + \hbar \omega_\perp + g_{1D}|f|^2\right]f
  \end{equation}
  where we defined $g_{1D} = g (N-1)/(2\pi l_\perp)$.
  The expression of $g_{1D}$ can be used to define a normalized nonlinear parameter $\gamma = (N-1) |a_s|/l_\perp$. Using Eq.~(\ref{eq:gpe}), for $g_{1D}<0$ the ground states of the axial problem are constituted by stable solitons, and no collapse is expected for any interaction strength. Instead, for the 3D-GPE case, there exists a critical nonlinear parameter for the existence of stable solitons. The value, above which static wave function collapse is expected, is about $\gamma_c \approx 0.67$.
  
\begin{figure} \centering
  \subfloat[Soliton solutions compared.]{\includegraphics[width=0.5\textwidth]{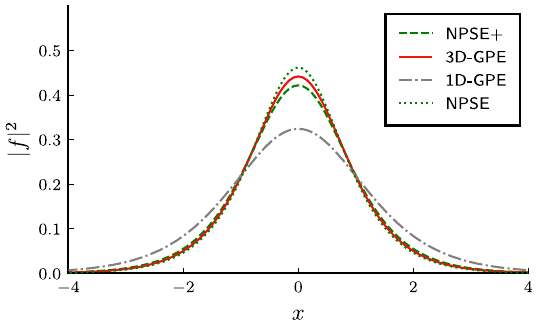}}
  \subfloat[Zoom applied to the top part of the soliton solutions.]{\includegraphics[width=0.5\textwidth]{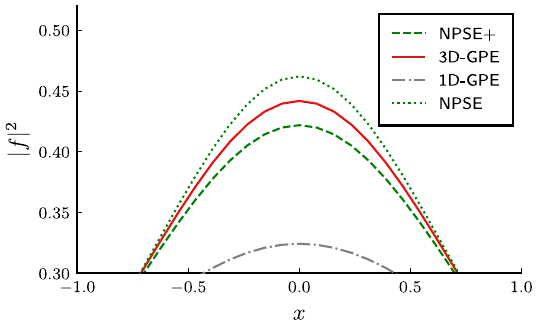}}
  \caption{Comparison of the axial ground state wave function $f$. The space coordinate $x$ is in units of $l_{\bot}=\sqrt{\hbar/(m\omega_{\bot})}$, the characteristic length of transverse harmonic confinement of frequency $\omega_{\bot}$. The nonlinear parameter is set to $\gamma=(N-1) |a_s|/l_\perp=0.65$. The three-dimensional Gross-Pitaevskii equation is the red solid line, the nonpolynomial Schr\"odinger equation without the corrections is the green dotted line, the one with the corrections is the green dashed line. The one-dimensional Gross-Pitaevskii equation {\cblue is the dash-dot grey line}.}
  \label{fig:gs}
\end{figure}

  \subsection*{Variational ansatz and NPSE}
  As shown in \cite{salasnich-effective-2002}, a better approximation is to consider the separation of the total wave function in a transverse Gaussian component with non-constant transverse width $\sigma(x, t)$ and to find the equation of motion using a variational principle. The resulting equation is the NPSE. The function $\phi$ in the ansatz Eq.~(\ref{eq:1dansats}) is substituted by a more general
  \begin{equation}
      \phi(y, z, \sigma(x, t)) = \dfrac{1}{\sqrt{\pi} \sigma(x, t)} \exp\left[-\frac{y^2+z^2}{2\sigma(x, t)^2}\right],
  \end{equation}
  where $\sigma$ is a function to be determined as a variational parameter.
  In \cite{salasnich-effective-2002}, the calculations were done assuming that derivatives of $\sigma$ are negligible.
  By keeping these derivatives terms, it is possible to write the following effective 1D Lagrangian (detailed calculations are given in the Methods section)
  \begin{equation}\label{eq:1dlagrangian_full}
          \mathcal{L}= \int \dd{x}  \ f^* \bigg[i \hbar \frac{\partial}{\partial t} + \frac{\hbar^2}{2 m} \frac{\partial^2}{\partial x^2} - V -\frac{\hbar^2}{2m} \frac{1}{\sigma^2}\left(1+ \left(\frac{\partial}{\partial x} \sigma\right)^2 \right) - \frac{m\omega_\perp^2}{2}\sigma^2 -\frac{\hbar^2 a_s (N-1)}{m \sigma^2}|f|^2 \bigg] f.
  \end{equation}
  The corresponding EL equations for $f$ and $\sigma$ provide the solution to the variational problem and are obtained as shown in \cite{salasnich-matter-wave-2007}:
  \begin{equation}\label{eq:ELf}
      i \hbar \frac{\partial}{\partial t} f = \bigg[- \frac{\hbar^2}{2 m} \frac{\partial^2}{\partial x^2} + V + \frac{\hbar^2}{2m} \frac{1}{\sigma^2}\left(1+ \left(\frac{\partial}{\partial x} \sigma\right)^2 \right) +\frac{m\omega_\perp^2}{2}\sigma^2 + \frac{2 \hbar^2 a_s (N-1)}{m \sigma^2}|f|^2 \bigg] f,
  \end{equation}
  \begin{equation}\label{eq:ELsigma}
          \sigma^4 - l_\perp^4\left[1 + 2a_s{\cplus (N-1)}|f|^2\right]
          +l_\perp^4\left[\sigma \frac{\partial^2}{\partial x^2} \sigma -\left(\frac{\partial}{\partial x}\sigma\right)^2 +\sigma \frac{\partial}{\partial x}\sigma \frac{1}{|f|^2}\frac{\partial}{\partial x} |f|^2 \right] = 0.
  \end{equation}
  We will refer to the above coupled equations as NPSE+.
  By neglecting the derivative of $\sigma$ in Eq.~(\ref{eq:1dlagrangian_full}), one obtains another effective 1D Lagrangian\cite{salasnich-effective-2002}, whose EL equations, called NPSE, correspond to
  \begin{equation}\label{eq:NPSE}
      \begin{split}
      i \hbar \frac{\partial}{\partial t} f = \bigg[- \frac{\hbar^2}{2 m} \frac{\partial^2}{\partial x^2} + V +\frac{\hbar^2}{2m} \frac{1}{\sigma^2} + \frac{m\omega_\perp^2}{2}\sigma^2 + \frac{2 \hbar^2 a_s (N-1)}{m \sigma^2}|f|^2 \bigg]f,
      \end{split}
  \end{equation}
  \begin{equation}\label{eq:NPSEsigma}
      \sigma^2 = l_\perp^2\sqrt{1+ 2a_s(N-1)|f|^2}.
  \end{equation}
  We remark that the NPSE+, as opposed to the NPSE, respects the variational principle, so the corresponding ground state energy is bound to be greater or equal to the true ground state energy of the 3D-GPE. 
  
  We will use the 3D-GPE as a reference equation, and compare the predictions of the 1D-GPE, the NPSE, and the NPSE+.
  The axial densities of the ground state solutions are shown in Fig.~\ref{fig:gs}, where we have set the nonlinear parameter to a very high value $\gamma=0.65$, near the 3D-GPE static collapse value $\gamma_c$. We notice that the 1D-GPE fails to represent accurately the axial wave function, NPSE+ and NPSE have similar accuracy. NPSE+ has the additional cost of the computation of the solution of the transverse width differential equation coupled to the axial wave function partial differential equation.

  \subsection*{Generalization of bound on splitting energy}
  The soliton splitting event can be verified only for specific ranges of the transmission coefficient \cite{gertjerenken-scattering-2012}, depending on the initial soliton velocity. These values can be computed by analyzing energy conservation in the splitting event. The interplay of kinetic energy and internal energy of the solitons during the scattering event has been discussed in \cite{helm-splitting-2014,wang-particle-wave-2012,gertjerenken-scattering-2012}, by using Lieb-Liniger\cite{lieb-exact-1963} energies $E_G$ pertaining to the soliton internal degrees of freedom in the total Hamiltonian. Imposing energy conservation, the kinetic energy of the initial soliton must satisfy 
  \begin{equation}
      E_k > E_G\left(N-n\right) + E_G\left(n\right) - E_G(N),
  \end{equation}
  where $N$ is the number of atoms in the initial soliton, and $n$ is the one in the transmitted soliton. This is the condition that must be satisfied for having a splitting event of transmission coefficient $T=n/N$.
  In our case, the internal energy of the soliton can be computed either numerically or analytically. The chemical potential of the stationary solution for the NPSE can be obtained \cite{salasnich-effective-2002} from the implicit relation
  \begin{equation}
    (1-\mu)^{3/2} - \dfrac{3}{2}(1-\mu)^{1/2} \dfrac{3}{2\sqrt{2}} \gamma= 0,
  \end{equation}
  and selecting only the stable branch of the solutions, i.e. the one satisfying the Vakhitov-Kolokolov criterion $\frac{\partial}{\partial n}\mu < 0$.
  For the other equations used in this work, it is possible to obtain numerically the value of $\mu(n)$ from stationary state solutions. 
  Using the values of the chemical potential, we are able to write the ground state energy of the nonlinear wave equation corresponding to an $N$-particle soliton as
  \begin{equation}
      E_G(n) = \int_0^n dn'\mu(n'),
  \end{equation}
  it is possible to obtain different ranges of transmission coefficients that are accessible for a given value of the initial kinetic energy.

\section{Results}\label{sec:num}
\begin{figure}
\begin{minipage}[c]{0.5\textwidth} \centering
  \includegraphics[width=\textwidth]{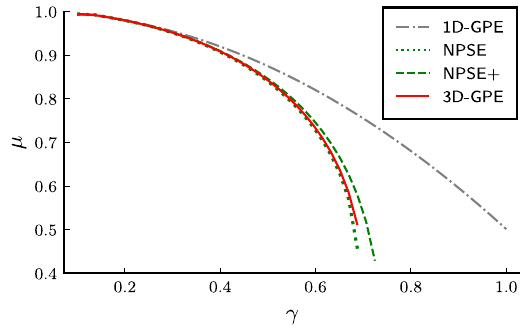}
  \caption{Chemical potential $\mu$ as a function of the nonlinear parameter $\gamma$. $\mu$ is in units of $\hbar\omega_{\bot}$.The three-dimensional Gross-Pitaevskii equation is the red solid line, the nonpolynomial Schr\"odinger equation without the corrections is the green dotted line, and the one with the corrections is the green dashed line. The one-dimensional Gross-Pitaevskii equation {\cblue is the dash-dot grey line}.}
  \label{fig:mu}
\end{minipage}%
\hspace{20pt}
\begin{minipage}[c]{0.5\textwidth} \centering
  \includegraphics[width=\textwidth]{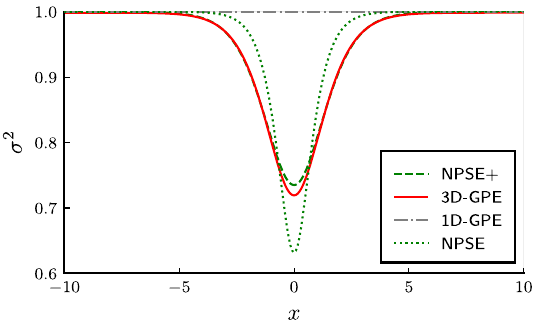}
  \caption{Comparison of the transverse width parameter $\sigma$. Both $\sigma$ and $x$ are in units of $l_{\bot}=\sqrt{\hbar/(m\omega_{\bot})}$. The colors are as in Fig.~\ref{fig:mu}.}
  \label{fig:sigma2}
\end{minipage}
\end{figure}

\begin{figure} \centering
  \subfloat[]{\includegraphics[width=0.5\textwidth]{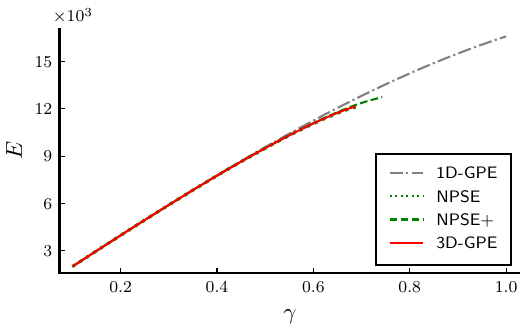}}
  \subfloat[Zoom applied to the region near static collapse.]{\includegraphics[width=0.5\textwidth]{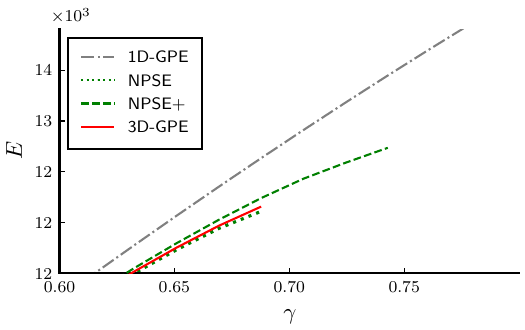}}
  \caption{
  Energy $E$ as a function of the nonlinear parameter $\gamma$, assuming a trap geometry such that $l_\perp / |a_s| = 2 \times 10^4$, that corresponds to a critical particle number $N_c \approx 13400$.  $E$ is in units of $\hbar \omega_\perp$. The three-dimensional Gross-Pitaevskii equation is the red solid line, the nonpolynomial Schr\"odinger equation without the corrections is the green dotted line, and the one with the corrections is the green dashed line. The one-dimensional Gross-Pitaevskii equation {\cblue is the dash-dot grey line}.}
  \label{fig:energy}
\end{figure}


\subsection*{Soliton solutions}
We review some properties of the solitonic ground state of the equations, comparing them.
We study the highly nonlinear regime, in which $\gamma=0.65$. Soliton solutions in this case are stable for the 3D-GPE and the NPSE and NPSE+ for $\gamma<\gamma_c$ \cite{salasnich-condensate-2002,sulem-nonlinear-2007}. 
Soliton solutions are shown for all the equations in Fig.~\ref{fig:gs} and their transverse width in Fig.~\ref{fig:sigma2}. The simulations show a better agreement of the NPSE+ equation with respect to the 3D-GPE in the transverse width. 
The computation of the NPSE+ transverse width is obtained by iteratively solving Eq.~(\ref{eq:ELf}) and then Eq.~(\ref{eq:ELsigma}). In the solution of the ordinary differential equation Eq.~(\ref{eq:ELsigma}) we apply Dirichlet boundary conditions corresponding to the vanishing of the axial wave function at infinity. 
The computation of the 3D-GPE transverse width is done by a least square fit on the radial distribution of the wave function, namely defined as
\begin{equation}
  \sigma^2(x) = \frac{1}{M(x)}\int\dd{y}\dd{z} \, (y^2+z^2)  |\psi(x,y,z)|^2,
\end{equation}
where 
\begin{equation}
    M(x) = \int \dd{y} \dd{z} |\psi(x, y, z)|^2.
\end{equation}
The chemical potential of the solitons is shown in Fig.~\ref{fig:mu}, and the corresponding energy is shown in Fig.~\ref{fig:energy}. We remark that, being the NPSE only an approximation of the true variational solution, its chemical potential is allowed to be less than the 3D-GPE chemical potential, thus becoming less than the bound set by the variational principle.
As expected, the difference becomes more pronounced at high nonlinearities, implying a more localized wave function, where the terms proportional to the derivatives of $\sigma$ in Eq.~(\ref{eq:ELf}) and (\ref{eq:ELsigma}) become more relevant.

\subsection*{Scattering from a narrow barrier}
\begin{figure*}
  \centering
    \includegraphics[width=.40\linewidth]{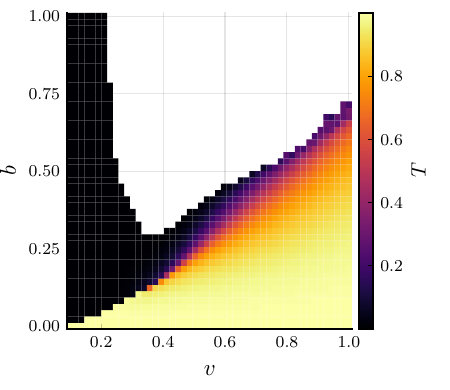}
  \caption{Transmission coefficient $T$ versus barrier height $b$ and velocity $v$, showing the collapse region for the 3D-GPE. $v$ is in units of $\sqrt{\hbar \omega_\perp / m}$, $b$ is in units of $\hbar \omega_\perp$.}
  \label{fig:heatmap}
\end{figure*}

\begin{figure} \centering
  \subfloat[$v=0.6$.]{\includegraphics[width=0.45\textwidth]{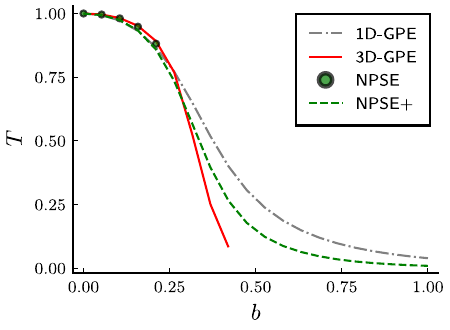}}
  \subfloat[$v=0.8$.]{\includegraphics[width=0.45\textwidth]{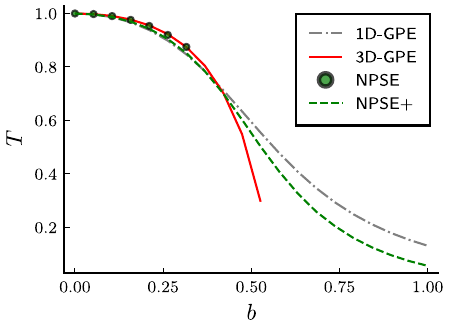}}
\caption{Transmission coefficient $T$ versus barrier height $b$ with fixed velocity $v$. $v$ that is in units of $\sqrt{\hbar \omega_\perp / m}$, and $b$ is in units of $\hbar \omega_\perp$. The three-dimensional Gross-Pitaevskii equation is the red solid line, the nonpolynomial Schr\"odinger equation without the corrections is the green dotted line, and the one with the corrections is the green dashed line. The one-dimensional Gross-Pitaevskii equation {\cblue is the dash-dot grey line}. }
\label{fig:transmission}
\end{figure}

We are interested in computing the transmission coefficient for various velocities and barriers.
We assume energy in units of $\hbar\omega_\perp$, time in units of $\omega_\perp^{-1}$, and length in units of $l_\perp$.
The barrier is Gaussian, centered in $x=0$, and it is parametrized by the peak value parameter $b$, 
\begin{equation}
    V(x; \ b) = b \ \exp[-\frac{x^2}{2 w^2}].
\end{equation}
The width $w$ is fixed to  $w=0.5 $.
In our simulations, the velocity ranges in $v\in [0.1, 1.0]$, and the barrier in $b \in [0.0, 1.0]$. In particular, by setting a sufficiently high $\gamma$, for example near to $\gamma_c$, namely $\gamma=0.65$, we analyze the onset of barrier-induced collapse, happening for high soliton velocity and high barrier height, as shown in Fig.~\ref{fig:heatmap}. Our results show that the vanishing of the transverse width predicted by the NPSE, suggesting a barrier-induced collapse, is a very weak indicator of an actual collapse. Instead, the NPSE+ collapsing region is not due to a vanishing of the {\cblue transverse width}, but to a sudden concentration of the axial density in smaller and smaller regions, like in the 3D-GPE case. In fact, we have set the numerical threshold of the collapse to the detection of a single probability per site greater than $0.3$. The abrupt change in the local maximum density we observe between stable solutions and collapsing ones justifies the validity of this criterion.  {\cblueA In the region of parameters we investigated, the NPSE+ is not collapsing, as reported in the transmission functions at constant velocity in Fig.~\ref{fig:transmission}, so it is ineffective in predicting the barrier-induced collapse present in the 3D-GPE. Results} obtained by the familiar 1D-GPE{\cblueA, are collapse-free.}
The comparison of the transmission coefficient versus the barrier height with fixed velocity reported in Fig.~\ref{fig:transmission} shows that, quite remarkably, the NPSE+ can describe accurately the transmission coefficient in the non-collapsing region.

\section{Discussion}\label{sec:conclusions}
In this article, we have presented a numerical study of the collision of a bright matter-wave soliton with a narrow potential barrier using the three-dimensional Gross-Pitaevskii and three dimensionally reduced versions of it. We investigated how the choice of dimensional reduction impacts the description of some features of the process, namely the transmission coefficient and the onset of barrier-induced collapse, also using the familiar one-dimensional Gross-Pitaevskii. We first reviewed the ground state properties given by all the schemes, highlighting the role of the variational transverse width. Then we compared the scattering properties: our results show that by using the NPSE in a regime of high barrier height and high velocity it fails to describe the 3D dynamics due to the the vanishing of the transverse width of the solution. In such cases the collapse phenomena in the 3D solutions are absent, and the NPSE is not capturing the correct dynamics. 

Our main result is that by adopting a slight modification of the NPSE, by using the true variational solution with the NPSE ansatz {\cblueA called NPSE+, we can predict the transmission factor and the dynamics of the transverse width more accurately, even though the collapse phenomenon is not captured by this effective equation.}
We believe the present work is a valuable contribution to the field of matter-wave soliton interferometry and quantum measurement, as results can be used to predict the dynamics of experimentally accessible scenarios. For example, the NPSE+ can be used for modeling interferometric experiments in highly nonlinear regimes where the determination of the transverse width is important. In the setting of a quasi 1D harmonic trap corresponding to a transverse frequency $\omega_\perp / 2\pi = 254 \text{Hz}$, loaded with about $28\times 10^3$ atoms of $^7$Li, analogously to a past experiment \cite{nguyen2014collisions}, the nonlinear regime we have studied is achieved for an s-wave scattering length of  $a_s \approx -5.52 \times 10^{-11} \text{m}$.

\section{Methods}\label{sec:appendix}

  \subsection*{Numerical methods}
  The time-marching scheme we use for all the simulations is the split-step Fourier method (SSFM), adopting Strang splitting of the nonlinear and linear part of the evolution operator. The SSFM is well-known to be accurate to the second order in time and to every order in space, thus being highly efficient in the spatial discretization \cite{taha-analytical-1984, taha-numerical-1984}. The drawback of the method - or the feature, depending on the point of view - is to natively implement periodic boundary conditions. The implementation of absorption boundaries in the context of this method is still possible but not straightforward \cite{antoine-computational-2013, antoine-perfectly-2020}. We assume the field to be localized away from the boundary in order to neglect this problem.
  In our setup, we use a unit of energy of $\hbar\omega_\perp$, a unit of time of $\omega_\perp^{-1}$ and a unit of length of $l_\perp$, constituting the natural units for the (isotropic) harmonic confinement. In these units, we consider for the 1D simulations a total length of $L = 40$, with a grid of $N = 512$ points. 
  In the 3D simulations, we use a grid of $(N_x, N_y, N_z) = (512, 40, 40)$ points, with  total lengths of $(L_x, L_y, L_z) = (40, 10, 10)$.
  The time step in both setups is chosen to be $h_t = 0.01$. These parameters have been proven to give a total truncation error in the $L_{\infty}$ norm of the order of $10^{-4}$ in 1D solitonic ground state solutions and 3D linear problems with anisotropic three-dimensional harmonic trap.
  
  The ground state solutions are computed using an imaginary-time propagation method. We point out that some modifications of this method are available under the name of normalized gradient-flow methods \cite{bao-computing-2004}.
\subsection*{Derivation of the NPSE+}
Following \cite{salasnich-condensate-2002}, we write the 3D Lagrangian
	\begin{equation}
        \mathcal{L}= \int \dd{x} \int\dd{y}\, \dd{z} \ f^* \phi^*\left[i \hbar \frac{\partial}{\partial t} + \right. \left.
        \frac{\hbar^2}{2 m} \nabla^2-U-\frac{1}{2} g(N-1)|f \phi|^2 \right] f \phi.
	\end{equation}
    We are interested in integrating along the transverse coordinates without neglecting the terms proportional to $\frac{\partial}{\partial x} \sigma$ and $\frac{\partial^2}{\partial x^2} \sigma$. The novel terms arise from $i\hbar\frac{\partial}{\partial t} (f\phi)$ and $\frac{\hbar}{2m}\nabla^2 (f\phi)$. By separating the derivatives, we have
    \begin{equation}
        \mathcal{L}= \int \dd{x} \int \dd{y}\dd{z} f^* \phi^*\biggl[i \hbar \phi \frac{\partial}{\partial t} f + i \hbar f  \phi \left(\dfrac{y^2+z^2}{\sigma^3} - \dfrac{1}{\sigma}\right)\frac{\partial}{\partial t} \sigma + 
         \frac{\hbar^2}{2 m} \left(f\nabla_\perp^2 \phi+ f \frac{\partial^2}{\partial x^2} \phi + \phi\frac{\partial^2}{\partial x^2} f \right)-U f \phi-\frac{1}{2} g(N-1)|f \phi|^2 f \phi\biggr].
    \end{equation}
    Integrating the term proportional to $\frac{\partial}{\partial t} \sigma$ gives $0$, as one may realize by looking at the symmetry of its prefactor. However, the term proportional to $\frac{\partial^2}{\partial x^2} \phi$ gives a non-null contribution to the 1D Lagrangian.
    The integration gives
    \begin{equation}
      \mathcal{L} = \int \dd{ x}\, f^* \bigg[ i \hbar \frac{\partial}{\partial t} + \frac{\hbar^2}{2m} \frac{\partial^2}{\partial x^2} - V - \frac{\hbar^2}{2m\sigma^2} \left(1 + \left(\frac{\partial}{\partial x}\sigma\right)^2\right) - \frac{m \omega_{\perp}^2}{2} \sigma^2 - \frac{1}{2} \frac{g(N-1)}{2\pi\sigma^2}|f|^2 \bigg] f.
      \end{equation}
      By considering the Euler-Lagrange equations, we recover Eq.~(\ref{eq:ELf}) and Eq.~(\ref{eq:ELsigma}).

\section*{Acknowledgements}

F.L. and L.S. acknowledge a National Grant of the Italian Ministry of 
University and Research for the PRIN 2022 project "Quantum Atomic Mixtures: Droplets, Topological Structures, and Vortices". L.S. is partially supported by the BIRD grant “Ultracold atoms in curved geometries” of the University of Padova, by the “Iniziativa Specifica Quantum” of INFN, by the European Quantum Flagship project PASQuanS 2, and by the European Union-NextGenerationEU within the National Center for HPC, Big Data and Quantum Computing (Project No. CN00000013, CN1 Spoke 10: “Quantum Computing”). 


\begin{thebibliography}{10}

\urlstyle{rm}
\expandafter\ifx\csname url\endcsname\relax
  \def\url#1{\texttt{#1}}\fi
\expandafter\ifx\csname urlprefix\endcsname\relax\def\urlprefix{URL }\fi
\expandafter\ifx\csname doiprefix\endcsname\relax\def\doiprefix{DOI: }\fi
\providecommand{\bibinfo}[2]{#2}
\providecommand{\eprint}[2][]{\url{#2}}

\bibitem{khaykovich-formation-2002}
\bibinfo{author}{Khaykovich, L.} \emph{et~al.}
\newblock \bibinfo{journal}{\bibinfo{title}{Formation of a matter-wave bright
  soliton}}.
\newblock {\emph{\JournalTitle{Science}}} \textbf{\bibinfo{volume}{296}},
  \bibinfo{pages}{1290--1293} (\bibinfo{year}{2002}).

\bibitem{weiss-creation-2009}
\bibinfo{author}{Weiss, C.} \& \bibinfo{author}{Castin, Y.}
\newblock \bibinfo{journal}{\bibinfo{title}{Creation and detection of a
  mesoscopic gas in a nonlocal quantum superposition}}.
\newblock {\emph{\JournalTitle{Phys. Rev. Lett.}}}
  \textbf{\bibinfo{volume}{102}}, \bibinfo{pages}{010403}
  (\bibinfo{year}{2009}).

\bibitem{streltsov-scattering-2009}
\bibinfo{author}{Streltsov, A.~I.}, \bibinfo{author}{Alon, O.~E.} \&
  \bibinfo{author}{Cederbaum, L.~S.}
\newblock \bibinfo{journal}{\bibinfo{title}{Scattering of an attractive
  Bose-Einstein condensate from a barrier: Formation of quantum superposition
  states}}.
\newblock {\emph{\JournalTitle{Phys. Rev. A}}} \textbf{\bibinfo{volume}{80}},
  \bibinfo{pages}{043616} (\bibinfo{year}{2009}).

\bibitem{helm-sagnac-2015}
\bibinfo{author}{Helm, J.~L.}, \bibinfo{author}{Cornish, S.~L.} \&
  \bibinfo{author}{Gardiner, S.~A.}
\newblock \bibinfo{journal}{\bibinfo{title}{Sagnac interferometry using bright
  matter-wave solitons}}.
\newblock {\emph{\JournalTitle{Phys. Rev. Lett.}}}
  \textbf{\bibinfo{volume}{114}}, \bibinfo{pages}{134101}
  (\bibinfo{year}{2015}).

\bibitem{dunningham-sub-shot-noise-limited-2004}
\bibinfo{author}{Dunningham, J.~A.} \& \bibinfo{author}{Burnett, K.}
\newblock \bibinfo{journal}{\bibinfo{title}{Sub-shot-noise-limited measurements
  with Bose-Einstein condensates}}.
\newblock {\emph{\JournalTitle{Phys. Rev. A}}} \textbf{\bibinfo{volume}{70}},
  \bibinfo{pages}{033601} (\bibinfo{year}{2004}).

\bibitem{dunningham-using-2006}
\bibinfo{author}{Dunningham, J.~A.}
\newblock \bibinfo{journal}{\bibinfo{title}{Using quantum theory to improve
  measurement precision}}.
\newblock {\emph{\JournalTitle{Contemp. Phys.}}} \textbf{\bibinfo{volume}{47}},
  \bibinfo{pages}{257--267} (\bibinfo{year}{2006}).

\bibitem{sun-mean-field-2014}
\bibinfo{author}{Sun, Z.-Y.}, \bibinfo{author}{Kevrekidis, P.~G.} \&
  \bibinfo{author}{Krüger, P.}
\newblock \bibinfo{journal}{\bibinfo{title}{Mean-field analog of the
  Hong-Ou-Mandel experiment with bright solitons}}.
\newblock {\emph{\JournalTitle{Phys. Rev. A}}} \textbf{\bibinfo{volume}{90}},
  \bibinfo{pages}{063612} (\bibinfo{year}{2014}).


\bibitem{lieb-exact-1963}
\bibinfo{author}{Lieb, E.~H.} \& \bibinfo{author}{Liniger, W.}
\newblock \bibinfo{journal}{\bibinfo{title}{Exact analysis of an interacting
  Bose gas. i. the general solution and the ground state}}.
\newblock {\emph{\JournalTitle{Phys. Rev.}}} \textbf{\bibinfo{volume}{130}},
  \bibinfo{pages}{1605--1616} (\bibinfo{year}{1963}).

\bibitem{cornish-formation-2006}
\bibinfo{author}{Cornish, S.~L.}, \bibinfo{author}{Thompson, S.~T.} \&
  \bibinfo{author}{Wieman, C.~E.}
\newblock \bibinfo{journal}{\bibinfo{title}{Formation of bright matter-wave
  solitons during the collapse of attractive Bose-Einstein condensates}}.
\newblock {\emph{\JournalTitle{Phys. Rev. Lett.}}}
  \textbf{\bibinfo{volume}{96}}, \bibinfo{pages}{170401}
  (\bibinfo{year}{2006}).

\bibitem{sulem-nonlinear-2007}
\bibinfo{author}{Sulem, C.} \& \bibinfo{author}{Sulem, P.-L.}
\newblock \emph{\bibinfo{title}{The Nonlinear Schrödinger Equation:
  Self-Focusing and Wave Collapse}}.

\bibitem{berge-wave-1998}
\bibinfo{author}{Bergé, L.}
\newblock \bibinfo{journal}{\bibinfo{title}{Wave collapse in physics:
  principles and applications to light and plasma waves}}.
\newblock {\emph{\JournalTitle{Phys. Rep.}}} \textbf{\bibinfo{volume}{303}},
  \bibinfo{pages}{259--370} (\bibinfo{year}{1998}).

\bibitem{mazzarella-collapse-2009}
\bibinfo{author}{Mazzarella, G.} \& \bibinfo{author}{Salasnich, L.}
\newblock \bibinfo{journal}{\bibinfo{title}{Collapse of triaxial bright
  solitons in atomic Bose–Einstein condensates}}.
\newblock {\emph{\JournalTitle{Phys. Lett. A}}} \textbf{\bibinfo{volume}{373}},
  \bibinfo{pages}{4434--4437} (\bibinfo{year}{2009}).

\bibitem{gammal-critical-2002}
\bibinfo{author}{Gammal, A.}, \bibinfo{author}{Tomio, L.} \&
  \bibinfo{author}{Frederico, T.}
\newblock \bibinfo{journal}{\bibinfo{title}{Critical numbers of attractive
  Bose-Einstein condensed atoms in asymmetric traps}}.
\newblock {\emph{\JournalTitle{Phys. Rev. A}}} \textbf{\bibinfo{volume}{66}},
  \bibinfo{pages}{043619} (\bibinfo{year}{2002}).

\bibitem{gertjerenken-generating-2013}
\bibinfo{author}{Gertjerenken, B.} \emph{et~al.}
\newblock \bibinfo{journal}{\bibinfo{title}{Generating mesoscopic Bell states
  via collisions of distinguishable quantum bright solitons}}.
\newblock {\emph{\JournalTitle{Phys. Rev. Lett.}}}
  \textbf{\bibinfo{volume}{111}}, \bibinfo{pages}{100406}
  (\bibinfo{year}{2013}).

\bibitem{khaykovich-deviation-2006}
\bibinfo{author}{Khaykovich, L.} \& \bibinfo{author}{Malomed, B.~A.}
\newblock \bibinfo{journal}{\bibinfo{title}{Deviation from one dimensionality
  in stationary properties and collisional dynamics of matter-wave solitons}}.
\newblock {\emph{\JournalTitle{Phys. Rev. A}}} \textbf{\bibinfo{volume}{74}},
  \bibinfo{pages}{023607} (\bibinfo{year}{2006}).

\bibitem{salasnich-condensate-2002}
\bibinfo{author}{Salasnich, L.}, \bibinfo{author}{Parola, A.} \&
  \bibinfo{author}{Reatto, L.}
\newblock \bibinfo{journal}{\bibinfo{title}{Condensate bright solitons under
  transverse confinement}}.
\newblock {\emph{\JournalTitle{Phys. Rev. A}}} \textbf{\bibinfo{volume}{66}},
  \bibinfo{pages}{043603} (\bibinfo{year}{2002}).

\bibitem{salasnich-effective-2002}
\bibinfo{author}{Salasnich, L.}, \bibinfo{author}{Parola, A.} \&
  \bibinfo{author}{Reatto, L.}
\newblock \bibinfo{journal}{\bibinfo{title}{Effective wave equations for the
  dynamics of cigar-shaped and disk-shaped Bose condensates}}.
\newblock {\emph{\JournalTitle{Phys. Rev. A}}} \textbf{\bibinfo{volume}{65}},
  \bibinfo{pages}{043614} (\bibinfo{year}{2002}).


\bibitem{cuevas-interactions-2013}
\bibinfo{author}{Cuevas, J.}, \bibinfo{author}{Kevrekidis, P.~G.},
  \bibinfo{author}{Malomed, B.~A.}, \bibinfo{author}{Dyke, P.} \&
  \bibinfo{author}{Hulet, R.~G.}
\newblock \bibinfo{journal}{\bibinfo{title}{Interactions of solitons with a
  gaussian barrier: splitting and recombination in quasi-one-dimensional and
  three-dimensional settings}}.
\newblock {\emph{\JournalTitle{New J. Phys.}}} \textbf{\bibinfo{volume}{15}},
  \bibinfo{pages}{063006} (\bibinfo{year}{2013}).

\bibitem{salasnich-matter-wave-2007}
\bibinfo{author}{Salasnich, L.}, \bibinfo{author}{Malomed, B.~A.} \&
  \bibinfo{author}{Toigo, F.}
\newblock \bibinfo{journal}{\bibinfo{title}{Matter-wave vortices in
  cigar-shaped and toroidal waveguides}}.
\newblock {\emph{\JournalTitle{Phys. Rev. A}}} \textbf{\bibinfo{volume}{76}},
  \bibinfo{pages}{063614} (\bibinfo{year}{2007}).

\bibitem{gertjerenken-scattering-2012}
\bibinfo{author}{Gertjerenken, B.}, \bibinfo{author}{Billam, T.~P.},
  \bibinfo{author}{Khaykovich, L.} \& \bibinfo{author}{Weiss, C.}
\newblock \bibinfo{journal}{\bibinfo{title}{Scattering bright solitons: Quantum
  versus mean-field behavior}}.
\newblock {\emph{\JournalTitle{Phys. Rev. A}}} \textbf{\bibinfo{volume}{86}},
  \bibinfo{pages}{033608} (\bibinfo{year}{2012}).

\bibitem{helm-splitting-2014}
\bibinfo{author}{Helm, J.~L.}, \bibinfo{author}{Rooney, S.~J.},
  \bibinfo{author}{Weiss, C.} \& \bibinfo{author}{Gardiner, S.~A.}
\newblock \bibinfo{journal}{\bibinfo{title}{Splitting bright matter-wave
  solitons on narrow potential barriers: Quantum to classical transition and
  applications to interferometry}}.
\newblock {\emph{\JournalTitle{Phys. Rev. A}}} \textbf{\bibinfo{volume}{89}},
  \bibinfo{pages}{033610} (\bibinfo{year}{2014}).

\bibitem{wang-particle-wave-2012}
\bibinfo{author}{Wang, C.-H.}, \bibinfo{author}{Hong, T.-M.},
  \bibinfo{author}{Lee, R.-K.} \& \bibinfo{author}{Wang, D.-W.}
\newblock \bibinfo{journal}{\bibinfo{title}{Particle-wave duality in quantum
  tunneling of a bright soliton}}.
\newblock {\emph{\JournalTitle{Opt. Expr.}}} \textbf{\bibinfo{volume}{20}},
  \bibinfo{pages}{22675--22682} (\bibinfo{year}{2012}).

\bibitem{nguyen2014collisions}
\bibinfo{author}{Nguyen, J.~H.}, \bibinfo{author}{Dyke, P.},
  \bibinfo{author}{Luo, D.}, \bibinfo{author}{Malomed, B.~A.} \&
  \bibinfo{author}{Hulet, R.~G.}
\newblock \bibinfo{journal}{\bibinfo{title}{Collisions of matter-wave
  solitons}}.
\newblock {\emph{\JournalTitle{Nature Physics}}} \textbf{\bibinfo{volume}{10}},
  \bibinfo{pages}{918--922} (\bibinfo{year}{2014}).

\bibitem{taha-analytical-1984}
\bibinfo{author}{Taha, T.~R.} \& \bibinfo{author}{Ablowitz, M.~J.}
\newblock \bibinfo{journal}{\bibinfo{title}{Analytical and numerical aspects of
  certain nonlinear evolution equations. i. analytical}}.
\newblock {\emph{\JournalTitle{J. Comput. Phys.}}}
  \textbf{\bibinfo{volume}{55}}, \bibinfo{pages}{192--202}
  (\bibinfo{year}{1984}).

\bibitem{taha-numerical-1984}
\bibinfo{author}{Taha, T.~R.} \& \bibinfo{author}{Ablowitz, M.~I.}
\newblock \bibinfo{journal}{\bibinfo{title}{Analytical and numerical aspects of
  certain nonlinear evolution equations. ii. numerical, nonlinear
  Schr{\"o}dinger equation}}.
\newblock {\emph{\JournalTitle{J. Comput. Phys.}}}
  \textbf{\bibinfo{volume}{55}}, \bibinfo{pages}{203--230}
  (\bibinfo{year}{1984}).

\bibitem{antoine-computational-2013}
\bibinfo{author}{Antoine, X.}, \bibinfo{author}{Bao, W.} \&
  \bibinfo{author}{Besse, C.}
\newblock \bibinfo{journal}{\bibinfo{title}{Computational methods for the
  dynamics of the nonlinear Schr{\"o}dinger/Gross-Pitaevskii equations}}.
\newblock {\emph{\JournalTitle{Comput. Phys. Comm.}}}
  \textbf{\bibinfo{volume}{184}}, \bibinfo{pages}{2621--2633}
  (\bibinfo{year}{2013}).

\bibitem{antoine-perfectly-2020}
\bibinfo{author}{Antoine, X.}, \bibinfo{author}{Geuzaine, C.} \&
  \bibinfo{author}{Tang, Q.}
\newblock \bibinfo{journal}{\bibinfo{title}{Perfectly matched layer for
  computing the dynamics of nonlinear Schr{\"o}dinger equations by
  pseudospectral methods. application to rotating Bose-Einstein condensates}}.
\newblock {\emph{\JournalTitle{Commun. Nonlinear. Sci. Numer. Simulat.}}}
  \textbf{\bibinfo{volume}{90}}, \bibinfo{pages}{105406}
  (\bibinfo{year}{2020}).

\bibitem{bao-computing-2004}
\bibinfo{author}{Bao, W.} \& \bibinfo{author}{Du, Q.}
\newblock \bibinfo{journal}{\bibinfo{title}{Computing the ground state solution
  of Bose-Einstein condensates by a normalized gradient flow}}.
\newblock {\emph{\JournalTitle{SIAM J. Sci. Comput.}}}
  \textbf{\bibinfo{volume}{25}}, \bibinfo{pages}{1674--1697}
  (\bibinfo{year}{2004}).

\bibitem{repo} Lorenzi, F. SolitonDynamics.jl \textit{GitHub repository} \texttt{github.com/lorenzifrancesco/SolitonDynamics.jl} (2023).
\end{thebibliography}

 \section*{Data availability statement}

The datasets used and/or analyzed during the current study are available from the corresponding author upon reasonable request. \\
The code developed for the current study is available at the public repository \cite{repo}.

\section*{ORCID IDs}
F Lorenzi https://orcid.org/0000-0001-8258-801X\\
L Salasnich https://orcid.org/0000-0003-0817-4753

\end{document}